\documentclass[12pt]{article}
\usepackage{epsfig}
\usepackage{graphicx}
\usepackage{amsfonts}
\usepackage{amsmath}
\usepackage{amssymb}
\usepackage{mathrsfs}
\usepackage{hhline,color}
\usepackage{pifont,ulem}
\usepackage{lmodern} 
\usepackage[utf8]{inputenc}

\newcommand{\pr}[1]{\ensuremath{\left[#1\right]}}
\newcommand{\pc}[1]{\ensuremath{\left(#1\right)}}
\newcommand{\chav}[1]{\ensuremath{\left\{#1\right\}}}
\newcommand{\ev}[1]{\ensuremath{\left\langle #1\right\rangle}}

\def\beq{\begin{equation}}
\def\eeq#1{\label{#1}\end{equation}}
\def\eeqn{\end{equation}}

\def\beqa{\begin{eqnarray}}
\def\eeqa#1{\label{#1}\end{eqnarray}}
\def\eeqan{\end{eqnarray}}

\let\bar=\overbar

\def\Dslash{\not{\hbox{\kern-4pt $D$}}}
\def\dslash{\not{\hbox{\kern-2pt $\del$}}}

\def\msb{{\bar{\ssstyle M \kern -1pt S}}}

\usepackage{fancyhdr,graphicx}
\def\Title#1{\begin{center} {\Large {\bf #1} } \end{center}}

\begin{document}

\Title{Inverse Magnetic Catalysis in hot quark matter within (P)NJL models}

\bigskip\bigskip

\begin{raggedright}

{\it 
M. Ferreira,$^{1,3}$ P. Costa,$^{1}$ C. Provid\^encia,$^{1}$
O. Louren\c co$^{\,2}$ and T. Frederico$^{\,3}$\\
\bigskip
$^{1}$Centro de F\'{\i}sica Computacional, 
Department of Physics, University of Coimbra,
P-3004 – 516 Coimbra, 
Portugal\\
\bigskip
$^{2}$Departamento de Ci\^encias da Natureza, 
Matem\'atica e Educa\c c\~ao, CCA, Universidade Federal de S\~ao Carlos,
13600-970 Araras, 
S\~ao Paulo, 
Brazil\\
\bigskip
$^{3}$Instituto Tecnol\'ogico de Aeron\'autica, 
12228-900 S\~ao Jos\'e dos Campos, 
S\~ao Paulo, 
Brazil\\
}

\end{raggedright}

\section{Magnetic Catalysis vs. Inverse Magnetic Catalysis in hot quark matter}

In recent years, lattice QCD calculations \cite{baliJHEP2012,bali2012PRD,endrodi2013}
and effective quarks models 
\cite{Menezes:2008qt.Menezes:2009uc,Ferreira:2013tba,Costa:2013zca,Ferreira:2014kpa,
Ferreira:2013oda,Ferreira:2014exa}
have been intensively used to investigate magnetized quark matter. 
An external magnetic field affects the QCD phase 
diagram structure based in the competition of two opposite mechanisms:
on the one hand the magnetic field enhances the chiral condensate due the opening 
of the gap between the Landau levels, increasing the low-energy 
contributions to the chiral condensate; 
on the other hand it contributes to the suppression of the condensate due to the 
strong screening effect of the gluon interactions in the region of the low momenta 
relevant for the chiral symmetry breaking mechanism \cite{Miransky:2002rp}. 
This suppression of the quark condensate, also known as inverse magnetic 
catalysis (IMC), manifests itself in the decreasing of the pseudocritical chiral 
transition temperature obtained in LQCD calculations with physical quark masses 
\cite{baliJHEP2012,bali2012PRD} and in the increasing of the Polyakov loop 
\cite{endrodi2013}.

In almost all effective quarks models, including the Nambu--Jona-Lasinio (NJL) 
model and the Polyakov--Nambu--Jona-Lasinio (PNJL) model \cite{PNJL}, with its 
generalizations like the Entangled PNJL (EPNJL) model \cite{Sakai:2010rp}, the 
inclusion of a magnetic field in the Lagrangian density allows describing the 
magnetic catalysis (MC) effect, i.e., the enhancement of the condensate due to 
the magnetic field, but fails to account for the IMC.  
In fact, for the NJL model the quarks interact through local 
current-current couplings, assuming that the gluonic degrees of freedom can be 
frozen into point like effective interactions between quarks. This leads to the
MC effect in the presence of an external magnetic field. 
Nevertheless, we may expect that the screening of the gluon interaction, discussed 
above, weakens the interaction which can be translated into a decrease of the 
scalar coupling with the intensity of the magnetic field. There are several recent 
studies that show a weakening effect of the coupling due to the magnetic field presence,
and that could be responsible for the of inverse magnetic catalysis mechanism 
\cite{Mueller:2015fka,Ayala:2014uua,Ayala:2014gwa,Ayala:2014iba}. 
Recently, two mechanisms were proposed within NJL-type models that can solve 
this discrepancy with implications in the structure of the QCD phase diagram:
\begin{itemize}
\item[-] by using the EPNJL \cite{Ferreira:2013tba} it was proposed that the 
parameter $T_0$ that enters in the Polyakov loop potential, that sets the transition 
temperature for pure-glue QCD lattice calculations \cite{PNJL}, depends on the 
magnetic field like it can depend on the number of quarks (and on the chemical 
potential at finite density);
\item[-] by using the SU(2) NJL model \cite{Farias:2014eca} and the SU(3) NJL/PNJL 
models \cite{Ferreira:2014kpa}, the model coupling, $G_s$, which can be seen as
proportional to the running coupling, $\alpha_s$, is made a decreasing function 
of the magnetic field strength allowing to include the impact of $\alpha_s$ in 
the models.
\end{itemize}

In the present work we will see the implications of the effective coupling 
that is a function of the magnetic field, $G_s (eB)$, on the quark condensates
and on the Polyakov loop, respectively the chiral and deconfinement order parameters. 
In order to do it, we will use the 2+1 PNJL model to describe quark matter 
subject to strong magnetic fields. The Lagrangian densities in the presence of 
an external magnetic field within this model is given by:
\begin{eqnarray}
{\cal L} = {\bar{q}} \left[i\gamma_\mu D^{\mu}-{\hat m}_f \right ] q ~+~ 
	G_s \sum_{a=0}^8 \left [({\bar q} \lambda_ a q)^2 + ({\bar q} i\gamma_5 \lambda_a q)^2 \right ]\nonumber\\
	-K\left\{{\rm det} \left [{\bar q}(1+\gamma_5) q \right] + 
	{\rm det}\left [{\bar q}(1-\gamma_5)q\right]\right\} + 
	\mathcal{U}\left(\Phi,\bar\Phi;T\right) - \frac{1}{4}F_{\mu \nu}F^{\mu \nu},
	\label{Pnjl}
\end{eqnarray}
where besides the chiral point-like coupling $G_s$, that denotes the coupling of 
the scalar-type four-quark interaction in the NJL sector, the quarks couple to a 
(spatially constant) temporal background gauge field, represented in terms of 
the Polyakov loop.
The Polyakov potential $\mathcal{U}\left(\Phi,\bar\Phi;T\right)$ is introduced 
and depends on the critical temperature $T_0$, that for pure gauge is 270 MeV 
but which we take as 210 MeV.

The thermodynamical potential for the three-flavor quark sector $\Omega$ is 
written as
\begin{align}
\Omega(T,\mu)&=G_s\sum_{f=u,d,s}\ev{\bar{q}_fq_f}^2
+4K\ev{\bar{q}_uq_u}\ev{\bar{q}_dq_d}\ev{\bar{q}_sq_s} 
+{\cal U}(\Phi,\bar{\Phi},T)\nonumber \\
&+\sum_{f=u,d,s}\pc{\Omega_{\text{vac}}^f+\Omega_{\text{med}}^f
+\Omega_{\text{mag}}^f}
\end{align}
where the flavor contributions from vacuum $\Omega^{\text{vac}}_f$, medium 
$\Omega^{\text{med}}_f$, and magnetic field $\Omega^{\text{mag}}_f$ 
\cite{Menezes:2008qt.Menezes:2009uc} are given by
\begin{align}
	\Omega_{\text{vac}}^f&=-6\int_{\Lambda}\frac{d^3p}{(2\pi)^3}E_f\\
	\Omega_{\text{med}}^f&=-T\frac{|q_fB|}{2\pi}\sum_{n=0}\alpha_n\int_{-\infty}^{+\infty}\frac{dp_z}{2\pi}\pc{Z_{\Phi}^+(E_f)+Z_{\Phi}^-(E_f)}\\
	\Omega_{\text{mag}}^f&=-\frac{3(|q_f|B)^2}{2\pi^2}\pr{\zeta^{'}(-1, x_f)-\frac{1}{2}(x_f^2-x_f)\ln x_f+\frac{x_f^2}{4}} 
\end{align}
where $E_f=\sqrt{p_z^2+M_f^2+2|q_f|Bk}$ , $\alpha_0=1$ and $\alpha_{k>0}=2$,
$x_f=M_f^2/(2|q_f|B)$, and $\zeta^{'}(-1, x_f)=d\zeta(z, x_f)/dz|_{z=-1}$, 
where $\zeta(z, x_f)$ is the Riemann-Hurwitz zeta function. 
At zero chemical potential the quark distribution functions 
$Z_{\Phi}^+(E_f)$ and $Z_{\Phi}^-(E_f)$ read
\begin{equation}
 Z_{\Phi}^+=Z_{\Phi}^-=\ln\chav{1+3\Phi e^{-\beta E_f}+3\Phi e^{-2\beta E_f}+e^{-3\beta E_f}}
\label{eq:z}
\end{equation}
once $\bar{\Phi}=\Phi$. 

\section{Inverse Magnetic Catalysis in the PNJL model}

As already mentioned, the strong coupling $\alpha_s$ should decrease 
with the the magnetic field strength. In the NJL model, the four-quark interaction
scalar coupling $G_s$, that can be seen as $\propto\alpha_s$,
must also be a decreasing function of $eB$.

Since there is no LQCD data available for $\alpha_s(eB)$, by using the NJL model 
we fit $G_s(eB)$ in order to reproduce the pseudocritical chiral transition 
temperatures, $T_c^\chi(eB)$, obtained in LQCD calculations \cite{baliJHEP2012}. 
The resulting fit function of $G_s(eB)$ that reproduces the $T_c^\chi(eB)$ 
is written as
\begin{equation}
G_s(\zeta)=G_s^0\pc{\frac{1+a\,\zeta^2+b\,\zeta^3}
{1+c\,\zeta^2+d\,\zeta^4}}\,
\label{eq:fit}
\end{equation}
with $a = 0.0108805$, $b=-1.0133\times10^{-4}$, $c= 0.02228$, and $d=1.84558\times10^{-4}$ and
where $\zeta=eB/\Lambda_{QCD}^2$. We have used $\Lambda_{QCD}=300$ MeV.

\begin{figure}[t]
\centering
    \includegraphics[width=0.45\linewidth,angle=0]{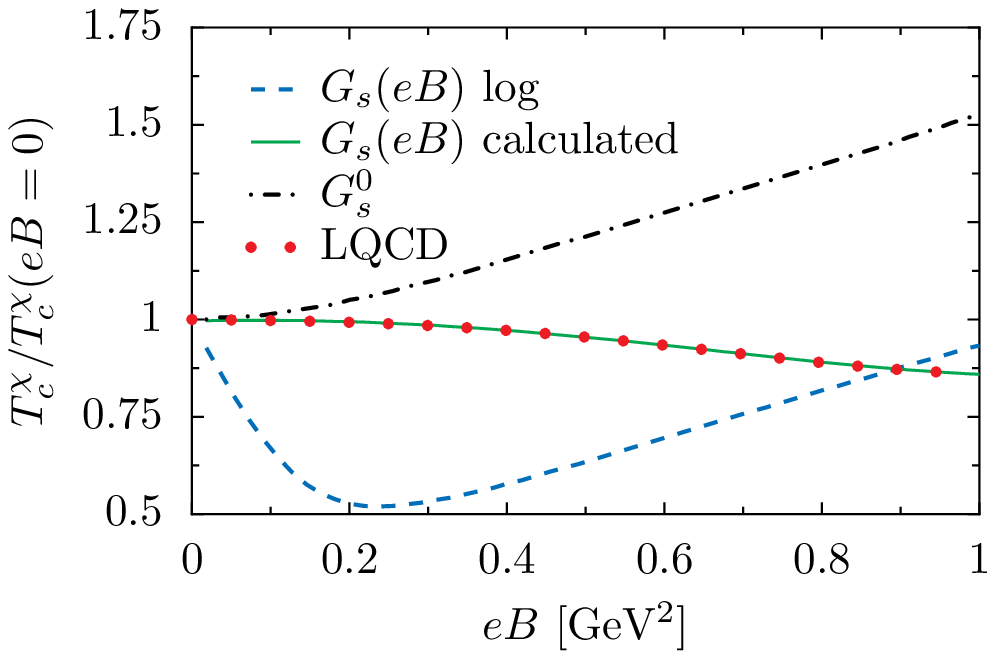}
		\includegraphics[width=0.45\linewidth,angle=0]{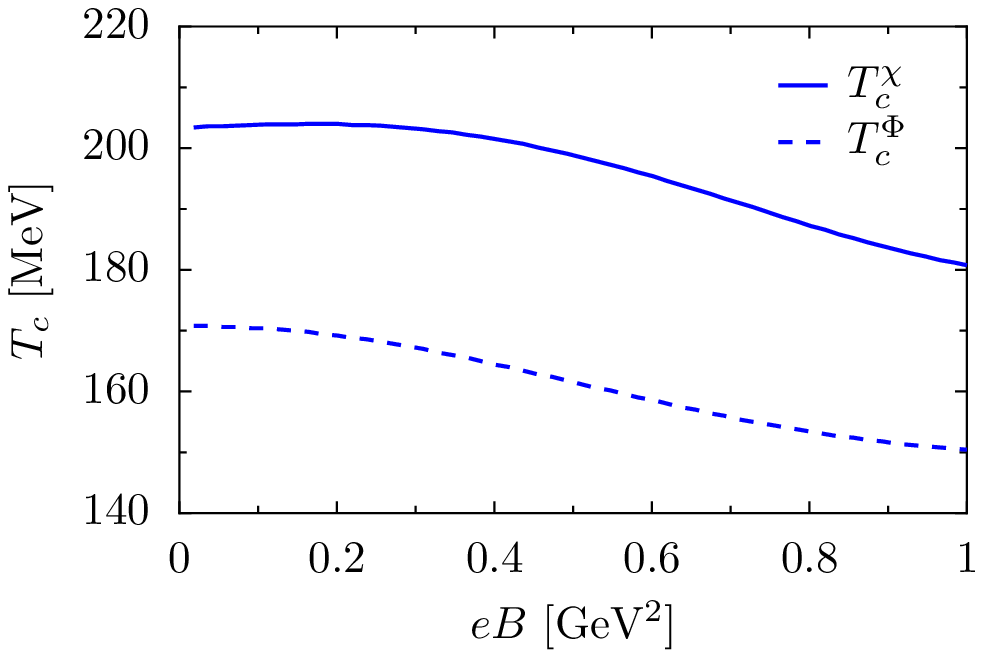}
\caption{
(Left panel) The renormalized critical temperatures
of the chiral transition ($T_c^\chi(eB=0)=178$ MeV) as a function of $eB$ in the NJL model with a
magnetic field dependent coupling $G_s(eB)$ (blue) and a constant
coupling $G_s^0$ (black), and the lattice results (red) \cite{baliJHEP2012}.
(Right panel) The chiral ($T_c^{\chi}$) and deconfinement ($T_c^{\Phi}$) transitions
temperatures as a function of $eB$ in the PNJL, using the magnetic
field dependent coupling $G_s(eB)$ [Eq. (\ref{eq:fit})].
}
\label{fig:temp_crit}
\end{figure}

The results for the renormalized critical temperature, $T_c^{\chi}/T_c^{\chi}(eB=0)$, 
of the pseudocritical chiral transition as a function of $eB$ in the NJL model, 
with the magnetic field dependent coupling $G_s(eB)$ given by Eq. (\ref{eq:fit}) 
is plotted in left panel of Fig. \ref{fig:temp_crit} (green line) together with 
LQCD results (red dots), the usual constant coupling $G_s=G_s^0$ (black dashed 
dot line) and the ansatz given by $G_s(eB)=\alpha_s(eB)=1/(b\ln|eB|\Lambda_{QCD}^2)$
with $b=(11N_c-2N_f)/12\pi=27/12\pi$ \cite{Miransky:2002rp} (blue dashed line). 
When $G_s=G_s^0$ the model always shows a magnetic catalyzes with increasing 
$T_c^{\chi}/T_c^{\chi}(eB=0)$ for all range of magnetic fields. 
If we consider $G_s(eB)=1/(b\ln|eB|\Lambda_{QCD}^2)$ \cite{Miransky:2002rp}, 
an IMC is seen until $eB \approx 0.3$ GeV$^2$, with the decrease of the 
pseudocritical temperature for this low magnetic fields.
However, for $eB  \gtrsim 0.3$ GeV$^2$, $T_c^{\chi}/T_c^{\chi}(eB=0)$ increases. 

Taking the magnetic field dependent coupling, $G_s(eB)$, given by Eq. (\ref{eq:fit}) 
we calculate the chiral and deconfinement transitions temperatures as a function 
of $eB$ in the PNJL model. The results are presented in the right panel of Fig. 
\ref{fig:temp_crit}: due to the coupling between the Polyakov loop field 
and quarks within the PNJL model, the $G_s(eB)$ does not only affect the chiral 
transition but also the deconfinement transition, so, both transitions temperatures 
decrease with the increase of the magnetic filed. 

\begin{figure}[h]
\centering
    \includegraphics[width=0.45\linewidth,angle=0]{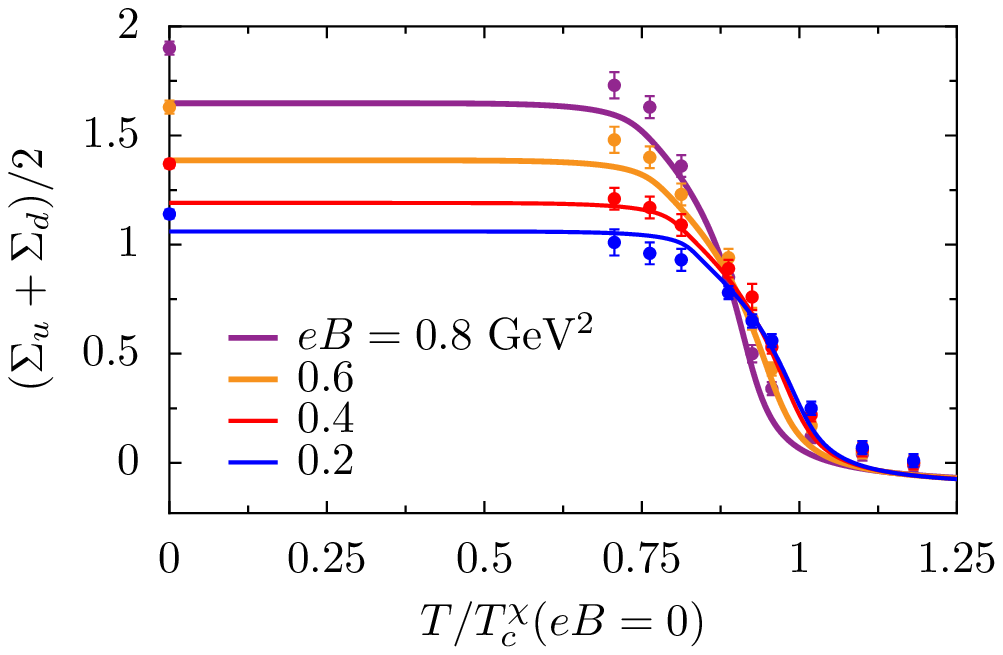}
		\includegraphics[width=0.45\linewidth,angle=0]{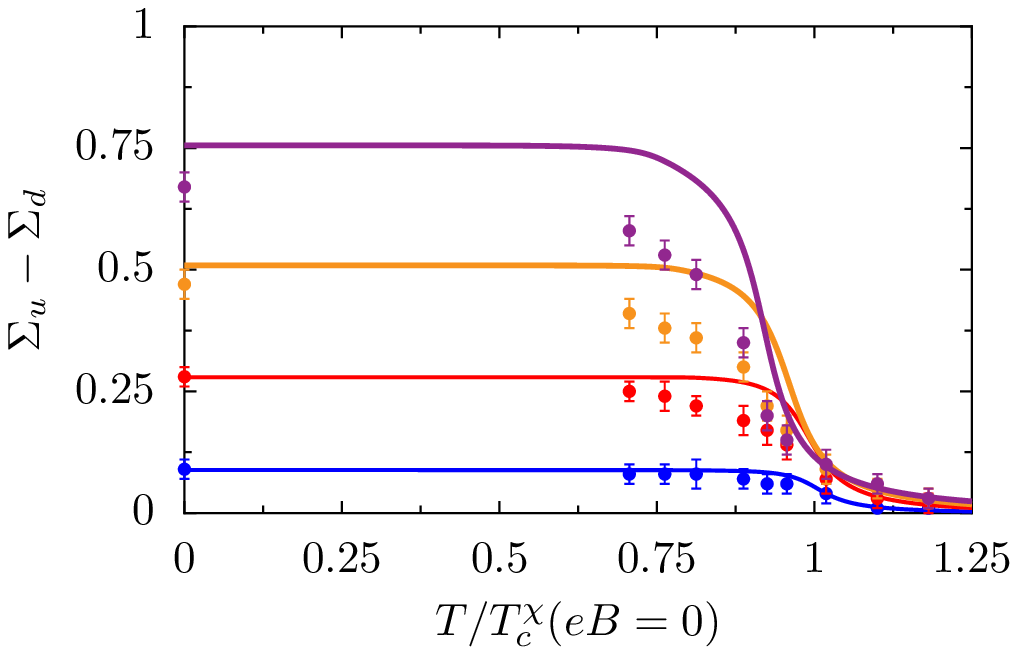}
\caption{
Average $(\Sigma_u+\Sigma_d)/2$ (left panel) and difference $(\Sigma_u-\Sigma_d)$ 
(right panel) of the light chiral condensates as a function of the renormalized temperature, 
for several values of $eB$, and LQCD results \cite{bali2012PRD}.
The LQCD data was renormalized by $T_c^\chi(eB=0)=160$ MeV \cite{bali2012PRD} and the PNJL model
results by $T_c^\chi(eB=0)=203$ MeV.
}
\label{fig:condensate}
\end{figure}

In Fig. \ref{fig:condensate} the results for the average chiral condensate, 
$(\Sigma_u+\Sigma_d)/2$, and the chiral condensate difference, 
$\Sigma_u-\Sigma_d$, are plotted as functions of $T_c^{\chi}/T_c^{\chi}(eB=0)$
for several magnetic field strengths and compared with the LQCD results from 
\cite{bali2012PRD}.  
We observe a qualitative agreement between both calculations for 
$(\Sigma_u+\Sigma_d)/2$ (left panel), meaning that the general features of the 
LQCD results are now reproduced.

\begin{figure}[h]
\centering
    \includegraphics[width=0.45\linewidth,angle=0]{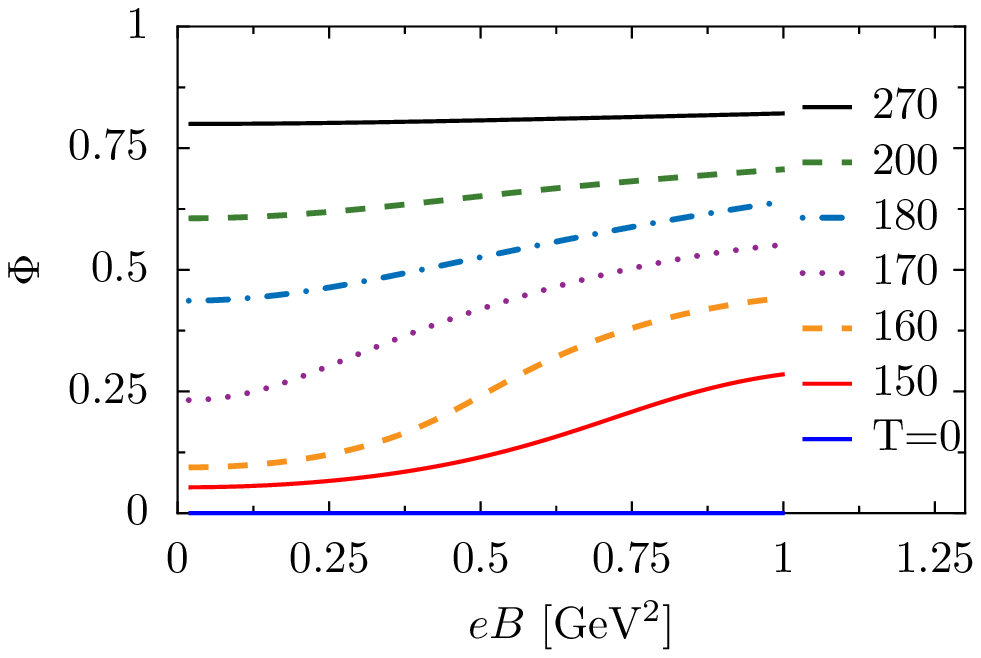}
		\includegraphics[width=0.45\linewidth,angle=0]{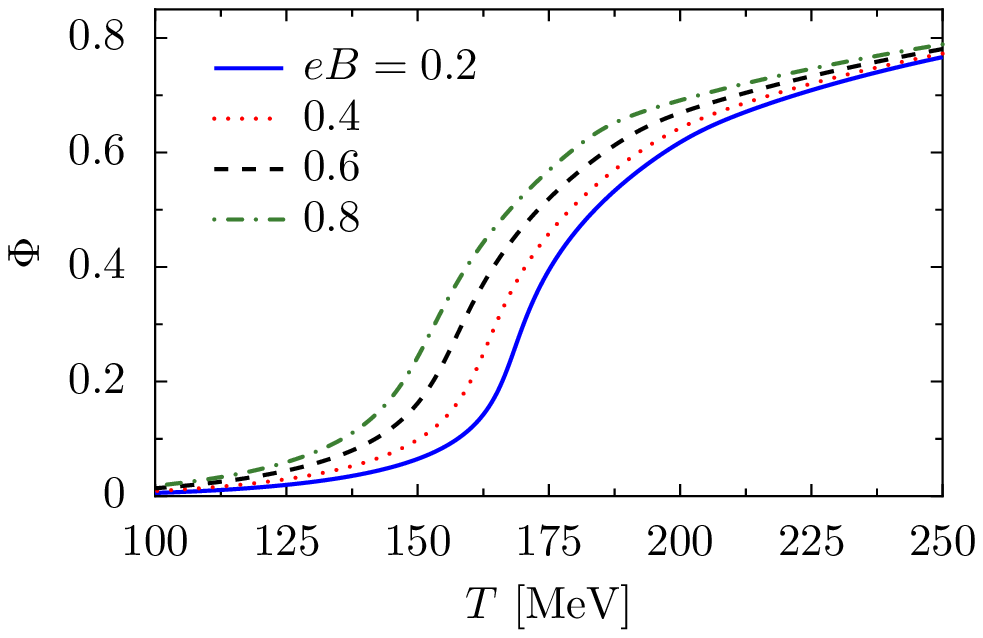}
\caption{
The value of the Polyakov loop versus $eB$ for several values of $T$ (MeV) 
(left panel) and versus $T$ for several values of $eB$ in GeV$^2$
(right panel).}
\label{fig:P_L}
\end{figure}

We also observe that SU(3) symmetry of the point like effective interactions 
between quarks is assumed in the magnetic background, however the comparison 
with the LQCD results for the difference in the quark condensates, 
$\Sigma_u-\Sigma_d$, in Fig. \ref{fig:condensate} right panel, suggests that 
the up quark interaction is depleted with respect to the down quark one. 
That, seems reasonable as the effect of the magnetic field on the up quark is 
larger than in the down quark, and, therefore, the interaction between the up 
quarks should decrease with respect to the down quarks as the magnetic field 
increases. 
Consequently, a more detailed calculation must also take into account that the SU(3) 
symmetry of the pointlike effective interaction between quarks should be
broken in the magnetic environment.

The effect of the magnetic field on the Polyakov loop is presented in Fig. 
\ref{fig:P_L} where $\Phi$ is plotted as a function of the magnetic field 
intensity for different values of the
temperature (left panel), and as a function of temperature, for several
magnetic field strengths (right panel). The suppression of the condensates
achieved by the magnetic field dependence of the coupling parameter is
translated in an increase of the Polyakov loop. The effect of the
magnetic field on $\Phi$ is stronger precisely for the temperatures close
to the transition temperature, see Fig. \ref{fig:P_L} (left panel), in
close agreement with the LQCD results \cite{endrodi2013}.

\subsection*{Acknowledgement}

This work was partially supported by Project No. PEst-OE/FIS/UI0405/2014
developed under the initiative QREN financed by the UE/FEDER through the
program COMPETE $-$ ``Programa Operacional Factores de
Competitividade'', and by Grant No. SFRH/BD/51717/2011.

\end{document}